\documentclass[%
 aip,
 pop,%
 amsmath,amssymb,
 reprint,%
]{revtex4-1}

\usepackage[normalem]{ulem}
\usepackage{color}
\usepackage{graphicx}
\usepackage{epstopdf}
\usepackage{dcolumn}
\usepackage{bm}
\usepackage{wasysym}

\begin{document}

\preprint{AIP/123-QED}

\title[Collision properties of small-amplitude supersolitons]
      {Collision properties of overtaking supersolitons with small amplitudes}

\author{C. P. Olivier}
\email{carel.olivier@nwu.ac.za}
\affiliation{
Centre for Space Research, North-West University, Potchefstroom 2520, South Africa}
\author{F. Verheest}%
 \email{frank.verheest@ugent.be}
\affiliation{
Sterrenkundig Observatorium, Universiteit Gent,
Krijgslaan 281, B--9000 Gent, Belgium
}
\affiliation{School of Chemistry and Physics, University of KwaZulu-Natal,
Durban 4000, South Africa}

\author{W. A. Hereman}
\email{whereman@mines.edu}
\affiliation{%
Department of Applied Mathematics and Statistics,
Colorado School of Mines, Golden, CO 80401-1887, USA
}%

\date{\today}

\begin{abstract}
The collision properties of overtaking small-amplitude supersolitons
are investigated for the fluid model of a plasma consisting of cold ions and
two-temperature Boltzmann electrons.
A reductive perturbation analysis is performed for compositional parameters near
the supercritical composition.
A generalized Korteweg-de Vries equation with a quartic nonlinearity is derived,
referred to as the modified Gardner equation.
Criteria for the existence of small-amplitude supersolitons
are derived.
The modified Gardner equation is shown to be not completely integrable,
implying that supersoliton collisions are inelastic, as confirmed by numerical simulations.
These simulations also show that supersolitons may reduce
to regular solitons as a result of overtaking collisions.
\end{abstract}

\maketitle

\section{Introduction}

Supersolitons are deformed solitary waves that are distinguishable
through their three local minima and three local maxima in the electric field.
Since the first reports on supersolitons, \cite{DubKol1,DubKol2,DubKol3}
an increasing number of plasma models that support
supersolitons have been identified.  \cite{VerhEtAl2013a,
VerhEtAl2013b,VerhEtAl2013c,MaharEtAl2013,OlivEtAl2015,
VerhOliv2017}
Many of these models describe magnetospheric plasmas.

Regardless, very few actual satellite observations of possible supersoliton profiles
have been reported.
\cite{VerhEtAl2013b,DubKol2018}
The limitations of spacecraft data means that the
time evolution of these structures cannot be traced.
It is therefore nearly impossible to distinguish
between supersolitons
and regular soliton collisions.

The observed supersoliton-like structures\cite{VerhEtAl2013b,DubKol2018} are typically
sandwiched  between regular solitons, or more complicated electric field structures.
This is not entirely unexpected,
as solitons in space plasmas are usually observed in clusters.\cite{TemEtAl1982,BoehmEtAl1984,BoundsEtAl1999,PickEtAl2009}
These observations suggest that supersolitons in space plasmas
would frequently collide with other solitons.
Therefore, it is important to understand the collision properties of supersolitons.

A fluid simulation was recently performed by
Kakad \textit{et al.} \cite{KakadEtAl2016} in order to investigate
the properties of supersolitons.
They simulated the formation of a supersoliton from a Gaussian initial
density disturbance.
The generated supersolitons are therefore stable
and provide insight into the possible formation of supersolitons.
However, the collision properties of the resulting supersolitons were not considered.

To date, theoretical studies have solely relied on pseudopotential
analysis due to Sagdeev. \cite{Sagd1966}
This approach is useful to obtain supersoliton solutions from which exact information
about their amplitudes, velocities, and parametric regions of existence
can be deduced.
Unfortunately, the study of collision properties falls outside the scope of Sagdeev analysis.

To study collision properties,
we will apply the reductive perturbation analysis of Washimi and Taniuti. \cite{WashTan1966}
Previously, it was suggested that reductive perturbation analysis cannot be used
to obtain small-amplitude supersolitons. \cite{VerhEtAl2013b,VerhHell2015}
But at that time, the existence of supercritical plasma compositions
\cite{VerhEtAl2016} had not been reported yet.
More recently, supercritical plasma compositions have been shown to be related to small-amplitude
supersolitons. \cite{OlivEtAl2017}

In this paper, we show how this relationship can be used to study
small-amplitude supersolitons by means of reductive perturbation analysis.
This requires an extension of the earlier reductive perturbation analysis
\cite{VerhEtAl2016} for a fluid plasma model consisting of cold ions and
two-temperature Boltzmann electrons.
The analysis leads to a generalized Korteweg-de Vries equation
that admits supersoliton solutions.
That equation is a higher order variant of the standard Gardner equation.
Since we have not come across this equation previously,
we refer to it as the modified Gardner (mG) equation.

The solutions obtained from this study agree exactly with those of an earlier
small-amplitude study based on Sagdeev potential analysis.\cite{OlivEtAl2017}
The main advantage of the reductive perturbation analysis is that one
may use the resulting evolution equation to analyze the collision
properties of the supersolitons in the small amplitude regime.
This is done in two ways. Firstly, we show that the
mG equation is not completely integrable. As a result,
it follows that supersoliton collisions are inelastic.
Secondly, we use the mG equation to simulate the
collision between solitons and overtaking supersolitons.
These simulations suggest that such collisions may reduce
the supersoliton to a regular soliton with smaller amplitude.

It should be noted that our study is limited to very small regions in parameter space,
very small amplitudes, and velocities that only marginally
exceed the acoustic speed. \cite{OlivEtAl2017}
Indeed, a comprehensive study of supersoliton collisions can only be undertaken
through full fluid simulations.
However, our results show that the collision properties of small-amplitude supersolitons
are very different from those of regular solitons.

The paper is organized as follows:
In Section 2 we present the fluid model.
In Section 3 we apply reductive perturbation analysis to derive the mG equation.
We also establish the necessary conditions for the existence of supersoliton solutions.
In Section 4 we normalize the mG equation and list its conservation laws.
Moreover, we discuss why the equation is not completely integrable.
Consequently, one should not expect collisions of solitons and supersolitons to be elastic.
In Section 5, we use the mG equation to simulate the collision of a supersoliton
that overtakes a regular soliton.
Some conclusions are drawn in Section 6 together with an outlook on future work.

\section{Fluid model}

We consider a plasma consisting of
cold fluid ions
and a two-temperature Boltzmann electron species.
The normalized fluid equations are given by\cite{VerhEtAl2016}
\begin{equation}
\frac{\partial n}{\partial t}+\frac{\partial}{\partial x}\left(nu\right)=0,
\label{eq:fluids 1}
\end{equation}
\begin{equation}
\frac{\partial u}{\partial t}+u\frac{\partial u}{\partial x}+\frac{\partial\phi}{\partial x}=0,
\label{eq:fluids 2}
\end{equation}
\begin{equation}
\frac{\partial^{2}\phi}{\partial x^{2}}+n-f\mbox{exp}\left(\alpha_{c}\phi\right)
-\left(1-f\right)\mbox{exp}\left(\alpha_{h}\phi\right)=0,
\label{eq:fluids 3}
\end{equation}
where $n$ denotes the ion number density normalized
with respect to the equilibrium ion density $N_{i}$, and $u$
the fluid velocity normalized with respect to the ion-acoustic speed
$c_{ia}=\sqrt{K_{\mathrm{B}}T_{\mathrm{eff}}/m_{i}}$ with ion mass $m_{i}$.
Here, $K_{\mathrm{B}}$ denotes the Boltzmann constant and $T_{\mathrm{eff}}=T_{c}/\left[f+\left(1-f\right)\sigma\right]$
denotes the effective temperature with electron temperature ratio
$\sigma=T_{c}/T_{h}$ for cool (hot, resp.) electron temperature
$T_{c}$ $\left(T_{h}, \mathrm{resp.}\right)$.
The cool electron density $f$ is normalized with respect to $N_{i}$. In addition, $\phi$ denotes the electrostatic
potential normalized with respect to $K_{\mathrm{B}}T_{\mathrm{eff}}/e$ where $e$
is the electron charge, while
\begin{equation}
\alpha_{c}=\frac{1}{f+\left(1-f\right)\sigma},
\end{equation}
and
\begin{equation}
\alpha_{h}=\frac{\sigma}{f+\left(1-f\right)\sigma}.
\end{equation}
Finally, length $x$ and time $t$ are normalized with respect to the Debye length
$\lambda_{\mathrm{D}}=\sqrt{\varepsilon_{0}\kappa T_{\mathrm{eff}}/\left(N_{i}e^{2}\right)}$
and the reciprocal of the plasma frequency, $\omega_{pi}^{-1}=\sqrt{\varepsilon_{0}m_{i}/N_{i}e^{2}}$,
respectively.

\section{Reductive perturbation analysis}

In order to retain fourth-order nonlinear effects, we follow Ref. \citenum{VerhEtAl2016}
and introduce a stretched coordinate system
\begin{equation}
\xi=\varepsilon^{3/2}\left(x-t\right), \quad \tau=\varepsilon^{9/2}t.
\label{eq:Stretching}
\end{equation}
In addition, we expand the ion number density and velocity, and the electrostatic potential as follows:
\begin{equation}
\left\{ \begin{array}{ccc}
n & = & 1+\varepsilon n_{1}+\varepsilon^{2} n_{2}+\varepsilon^{3} n_{3}+\varepsilon^{4} n_{4}+\cdots, \\
\\
u & = & \varepsilon u_1 + \varepsilon^2 u_2 +  \varepsilon^3 u_3 + \varepsilon^4 u_4 + \cdots,\\
\\
\phi & = & \varepsilon \phi_1 + \varepsilon^2 \phi_2 + \varepsilon^3 \phi_3 + \varepsilon^4 \phi_4 + \cdots .
\end{array} \right.
\label{eq:expansion}
\end{equation}
Since we are interested in solitons and supersolitons, we impose the following boundary conditions:
\begin{equation}
n\rightarrow1,u\rightarrow0,\:\phi\rightarrow0\text{ when }|\xi|\rightarrow\infty.
\label{eq:BCs}
\end{equation}
By substituting the expressions (\ref{eq:Stretching}) and (\ref{eq:expansion})
into the fluid equations (\ref{eq:fluids 1})--(\ref{eq:fluids 3}),
one obtains differential equations at different orders of $\varepsilon$.
For brevity, we do not present these long expressions.

We start with the continuity equation (\ref{eq:fluids 1}). By substituting the expansions (\ref{eq:Stretching}) and (\ref{eq:expansion}) into
the continuity equation, and collecting terms up to $\varepsilon^{11/2}$, one obtains the following equations:
\begin{equation}
n_{1\xi} = u_{1\xi}.\label{eq:Continuity expansion 1}
\end{equation}
\begin{equation}
n_{2\xi}=u_{2\xi}+\left(n_1u_1\right)_\xi.\label{eq:Continuity expansion 2}\\
\end{equation}
\begin{equation}
n_{3\xi}=u_{3\xi}+\left(n_1u_2+n_2u_1\right)_\xi.\label{eq:Continuity expansion 3}\\
\end{equation}
\begin{equation}
n_{4\xi}=n_{1\tau}+u_{4\xi}+\left(n_1u_3+n_2u_2+n_3u_1\right)_\xi.\label{eq:Continuity expansion 4}
\end{equation}
The subscripts $\xi$ and $\tau$ are used to denote partial derivatives $\partial/\partial\xi$ and $\partial/\partial\tau$, respectively.
In addition, higher order
partial derivatives are denoted with multiple subscripts throughout the paper. For example,
we use
$\phi_{1\xi\xi}$ to denote $\partial^2\phi_1/\partial\xi^2$.

The first three equations (\ref{eq:Continuity expansion 1})--(\ref{eq:Continuity expansion 3})
can be simplified by means of a simple integration. By taking the boundary conditions (\ref{eq:BCs})
into account, it follows that
\begin{equation}
n_1=u_1,\label{eq:Continuity simplified 1}
\end{equation}
\begin{equation}
n_2=u_2+n_1u_1,
\label{eq:Continuity simplified 2}\\
\end{equation}
and
\begin{equation}
n_3=u_3+n_1u_2+n_2u_1.\label{eq:Continuity simplified 3}\\
\end{equation}

A similar treatment of the momentum equation (\ref{eq:fluids 2})
produces the following set of equations:
\begin{equation}
u_1 = \phi_1,
\label{eq:Momentum simplified 1}
\end{equation}
\begin{equation}
u_2=\phi_2+\frac{1}{2}u_1^2,
\label{eq:Momentum simplified 2}
\end{equation}
\begin{equation}
u_3 = \phi_3+u_1u_2,
\label{eq:Momentum simplified 3}
\end{equation}
and
\begin{equation}
u_{4\xi}=u_{1\tau}+\phi_{4\xi}+u_1u_{3\xi}+u_2u_{2\xi}+u_3u_{1\xi}.
\label{eq:Momentum simplified 4}
\end{equation}

The set of equations (\ref{eq:Momentum simplified 1})--(\ref{eq:Momentum simplified 4}) can be combined to eliminate
the $u$ dependence from the set of equations (\ref{eq:Continuity expansion 4})--(\ref{eq:Continuity simplified 3}).
It follows that
\begin{equation}
n_1 = \phi_1,
\label{eq:Density order 1}
\end{equation}
\begin{equation}
n_2 = \phi_2 + \frac{3}{2}\phi_1^2,
\label{eq:Density order 2}
\end{equation}
\begin{equation}
n_3 = \phi_3 + 3\phi_1\phi_2+\frac{5}{2}\phi_1^3,
\label{eq:Density order 3}
\end{equation}
and 
\begin{equation}
n_{4\xi} = \phi_{4\xi}+2\phi_{1\tau}+\left(2\phi_1\phi_3+\frac{1}{2}\phi_2^2+\frac{3}{2}\phi_1^2\phi_2+\frac{5}{8}\phi_1^4\right)_\xi.
\label{eq:Density order 4}
\end{equation}

We now turn to Poisson's equation (\ref{eq:fluids 3}). By applying the expansions (\ref{eq:Stretching})
and (\ref{eq:expansion}), using a Taylor series to expand the exponential
functions, and retaining terms up to order $\varepsilon^4$, one obtains the following
equation:
\begin{equation}
\begin{array}{ccc}
\displaystyle{\varepsilon^4\phi_{1\xi\xi}+\varepsilon n_1+\varepsilon^2n_2+\varepsilon^3n_3+\varepsilon^4n_4-\varepsilon A_1\phi_1-\varepsilon^2A_1\phi_2}\\
\\
\displaystyle{-\varepsilon^3A_1\phi_3-\varepsilon^4A_1\phi_4-\frac{A_2}{2}\varepsilon^2\phi_1^2-A_2\varepsilon^3\phi_1\phi_2-A_2\varepsilon^4\phi_1\phi_3}\\
\\
\displaystyle{-\frac{A_2}{2}\varepsilon^4\phi_2^2-\frac{A_3}{6}\varepsilon^3\phi_1^3-\frac{A_3}{2}\varepsilon^4\phi_1^2\phi_2-\frac{A_4}{24}\varepsilon^4\phi_1^4=0,\qquad\qquad}
\end{array}
\label{eq:Poisson expanded}
\end{equation}
where
\begin{equation}
A_j = f\alpha_c^j + \left(1-f\right)\alpha_h^j.
\end{equation}

The equations (\ref{eq:Density order 1})--(\ref{eq:Density order 3}) must be substituted into (\ref{eq:Poisson expanded}).
To use (\ref{eq:Density order 4}), we differentiate (\ref{eq:Poisson expanded}) with respect to $\xi$. Since $A_1=1$ for
any choice of $f$ and $\sigma$, (\ref{eq:Poisson expanded}) becomes
\begin{equation}
\begin{array}{ccc}
\displaystyle{\varepsilon^4\phi_{1\xi\xi\xi}+2\varepsilon^4\phi_{1\tau}+\left[\frac{3-A_2}{2}\varepsilon^2\phi_1^2+\left(3-A_2\right)\varepsilon^3\phi_1\phi_2\qquad\right.}\\
\\
\displaystyle{+\frac{15-A_3}{6}\varepsilon^3\phi_1^3+\left(3-A_2\right)\varepsilon^4\phi_1\phi_3+\frac{15-A_3}{2}\varepsilon^4\phi_1^2\phi_2\qquad\qquad}\\
\\
\displaystyle{\left.+\frac{3-A_2}{2}\varepsilon^4\phi_2^2+\frac{105-A_4}{24}\varepsilon^4\phi_1^4\right]_\xi=0.\qquad\qquad\qquad\qquad\qquad}
\end{array}
\label{eq:Poisson expanded 2}
\end{equation}

For the supercritical plasma composition $f=\frac{1}{6}\left(3-\sqrt{6}\right)$ and $\sigma=5-2\sqrt{6}$,
one has $A_2=3$ and $A_3=15$, so that the terms
in orders $\varepsilon^2$ and $\varepsilon^3$ in (\ref{eq:Poisson expanded 2})
vanish.
Here we consider plasma compositions near the supercritical composition. To do so, we look
for compositions that satisfy the following criteria
\begin{equation}
A_2 = 3-\varepsilon^2 B_2,\,\,\,\,\,A_3 = 15-\varepsilon B_3.
\label{eq:Near supercritical comps}
\end{equation}
We thus require that $A_2$ is close to $3$ up to order $\varepsilon^2$ and
that $A_3$ only differs from $15$ by a quantity of order $\varepsilon.$
Obviously, $B_2$ and $B_3$ must both be of order $1.$

If we substitute (\ref{eq:Near supercritical comps}) into (\ref{eq:Poisson expanded 2}),
and retain terms of order $\varepsilon^4$,
we obtain
the following equation:
\begin{equation}
\phi_{1\tau}+\frac{1}{2}\phi_{1\xi\xi\xi}+\frac{B_2}{2}\phi_1\phi_{1\xi}+\frac{B_3}{4}\phi_1^2\phi_{1\xi}+\frac{105-A_4}{12}\phi_1^3\phi_{1\xi}=0.
\label{eq:mG 1}
\end{equation}

For further analysis of (\ref{eq:mG 1}), we consider the lowest order
approximation of the electrostatic potential
\begin{equation}
\Phi=\varepsilon\phi_1.
\end{equation}
In addition, we introduce the following changes of coordinates:
\begin{equation}
t=\varepsilon^{-9/2}\tau,\,\,\,\eta=\varepsilon^{-3/2}\xi=x-t.
\end{equation}
Then (\ref{eq:mG 1}) becomes
\begin{equation}
\Phi_{t}+\frac{1}{2}\Phi_{\eta\eta\eta}+a\Phi\Phi_{\eta}+b\Phi^2\Phi_{\eta}+c\Phi^3\Phi_{\eta}=0,
\label{eq:mG}
\end{equation}
where
\begin{equation}
a=\frac{3-A_2}{2},\,\,\,\,b=\frac{15-A_3}{4},\,\,\,\,c=\frac{105-A_4}{12}.
\label{eq:mG coefficients}
\end{equation}
To the best of our knowledge, (\ref{eq:mG}) has not been reported before in the literature.
We will refer to it as the modified Gardner (mG) equation
since it is a quartic version of the standard Gardner equation where $c=0$.

To find solitary wave solutions, we introduce a moving frame,
\begin{equation}
\zeta = \eta-vt,
\end{equation}
and integrate the resulting ordinary differential equation twice, to obtain
the energy-like equation,
\begin{equation}
\frac{1}{2}\frac{\partial}{\partial\zeta}\left(\Phi^2\right)+V\left(\Phi\right)=0,
\label{eq:Energy eq}
\end{equation}
where
\begin{equation}
V\left(\Phi\right)= -v\Phi^2+\frac{a}{3}\Phi^3+\frac{b}{6}\Phi^4+\frac{c}{10}\Phi^5.
\end{equation}
Note that the above Sagdeev potential
$V\left(\Phi\right)$ agrees with the one obtained in a small-amplitude study
\cite{OlivEtAl2017} based on a Taylor series expansion of the Sagdeev potential.
We briefly summarize the main results from that paper:
\begin{enumerate}
\item For the model under consideration, a supercritical plasma composition exists for $\sigma=5-2\sqrt{6}$ and $f=\left(3-\sqrt{6}\right)/6,$
yielding $A_2 = 3$ and $A_3 = 15,$
Using (\ref{eq:mG coefficients}),
it follows that $a = b = 0$ in (\ref{eq:mG}).

\item For supersolitons to exist,
the following conditions must be satisfied:
\begin{equation}
b<0,\qquad ac>0,\qquad ac<\frac{8}{27}b^2.
\label{eq:Existence crit}
\end{equation}

\item
For a plasma that satisfies these criteria, supersolitons exist at velocities
\begin{equation} v_{\mathrm{min}}<v<v_{\mathrm{max}}, 
\end{equation}
where
\begin{equation}
v_{\mathrm{max}}=v_{+},
\end{equation}
\begin{equation}
v_{\mathrm{min}} =
\left\{ \begin{array}{c}
v_{DL}\quad\text{ if }\quad\quad\quad\quad{\displaystyle \frac{ac}{b^{2}}\leq\frac{5}{18}},\\
\\
v_{-}\quad\text{ if }\quad{\displaystyle \frac{5}{18}<\frac{ac}{b^{2}}<\frac{8}{27}},
\end{array}\right.
\end{equation}
\begin{equation}
v_{DL}=\frac{5b{\displaystyle \left(\frac{5b^{2}-27ac}{27}\right)}-200{\displaystyle \left(\frac{5b^{2}-18ac}{180}\right)}^{3/2}}{27c^{2}},
\label{eq:DL}
\end{equation}
and
\begin{equation}
v_{\pm}=\frac{{\displaystyle \frac{2b}{27}}\left(16b^{2}-81ac\right)\pm4\left({\displaystyle \frac{8b^{2}-27ac}{18}}\right)^{3/2}}{27c^{2}}.
\end{equation}
In (\ref{eq:DL}), $v_{DL}$ corresponds to the velocity of a double layer solution.
\item A comparison between the small-amplitude study and the
analysis based on the fully nonlinear Sagdeev potential was performed.
Based on that comparison, a region (in the compositional parameter space) for the existence of small-amplitude supersolitons was found.
This region was established
for plasma compositions very close to the supercritical plasma composition.
\end{enumerate}
The mG equation can now be used to study collisions of overtaking
supersolitons of small amplitudes.

\section{Non-integrability of the mG equation}
Some of the coefficients in (\ref{eq:mG}) can be removed by scaling:
\begin{equation}
t\rightarrow\alpha t,\text{ }\eta\rightarrow\beta\eta,\text{ }\Phi\rightarrow\gamma\Phi.
\end{equation}
By choosing the parameters $\alpha, \beta,$ and $\gamma$ appropriately,
one obtains a normalized equation,
\begin{equation}
\Phi_{t} + \frac{1}{2}\Phi_{\eta\eta\eta}  \pm\Phi\Phi_{\eta} + d\Phi^2\Phi_\eta \pm\Phi^3\Phi_{\eta} = 0.
\label{eq:mG normalized}
\end{equation}
The signs in (\ref{eq:mG normalized}) and the choices of $\alpha$, $\beta$ and $\gamma$ depend
on the signs of $a$, $b$ and $c$.

For the model under consideration, one can easily show \cite{VerhEtAl2016} that
$A_4=81$ at the supercritical composition, so that $c=2$.
It can also easily be verified that $c>0$ for plasma compositions near the supercritical composition.
Based on the existence critera (\ref{eq:Existence crit}),
we restrict ourselves to compositions where $a>0$, $b<0$ and $c>0$.
Choosing the coefficients
\begin{equation}
\alpha=\left(\frac{c}{a}\right)^{3/4},\text{ }\beta=\left(\frac{c}{a}\right)^{1/4}, \text{ }\gamma=-\sqrt{\frac{a}{c}},
\end{equation}
yields
\begin{equation}
\Phi_{t} + \frac{1}{2}\Phi_{\eta\eta\eta} + \Phi\Phi_{\eta} + D\Phi^2\Phi_\eta + \Phi^3\Phi_{\eta} = 0,
\label{eq:mG normalized 2}
\end{equation}
with
\begin{equation}
D=-\sqrt{\frac{b^2}{ac}}=-\sqrt{\frac{3\left(15-A_3\right)^2}{2\left(3-A_2\right)\left(105-A_4\right)}}.
\end{equation}
To compute conservation laws of
(\ref{eq:mG normalized 2}), we follow the approach of Verheest and Hereman \cite{VerhHer1994}
which yields two conservation laws:
\begin{equation}
\Phi_t + \left( \frac{1}{2}\Phi^2 + \frac{D}{3} \Phi^3 + \frac{1}{4}\Phi^4 + \frac{1}{2} \Phi_{\eta\eta} \right)_\eta = 0,
\end{equation}
and
\begin{equation}
\left(\Phi^2\right)_t + \left( \frac{2}{3}\Phi^3 + \frac{D}{2} \Phi^4 + \frac{2}{5}\Phi^5 + \Phi\Phi_{\eta\eta} - \frac{1}{2} \Phi_\eta^2 \right)_\eta = 0.
\end{equation}
Using symbolic software developed by Poole and Hereman, \cite{PooleHereman2011} an extensive search for polynomial conservation laws of (\ref{eq:mG normalized 2})
did not yield any additional results which suggests that (\ref{eq:mG normalized 2}) is not completely integrable.
Equation (\ref{eq:mG normalized 2}) does not pass the Painlev\'e integrability test either as confirmed with the code of
Baldwin and Hereman.\cite{BaldwinHereman2006}
One should therefore not expect that solitary wave solutions of (\ref{eq:mG normalized 2}) would collide elastically and thus
retain their shapes upon collisions.

\section{Simulation of small-amplitude supersoliton collisions}

We can now use the mG equation to simulate collisions between solitons and supersolitons.
To do this, we construct a supersoliton solution and a slower soliton by
numerically integrating the energy equation (\ref{eq:Energy eq}).
The faster supersoliton solution is then shifted $\eta_0$ units to the left and added to the soliton solution.
While the principle of superposition does not apply to nonlinear equations,
it is assumed that the stability of the solutions ensures that the soliton and supersoliton
propagation remains unaffected provided that the two solutions are sufficiently
far apart.

It should be mentioned that the solutions must be constructed on a sufficiently large interval.
Due to the instability of the energy integral (\ref{eq:Energy eq}), the numerical integration is not
accurate enough to provide such solutions.
We therefore applied the results from an asymptotic study \cite{OlivEtAl2017_2} to construct sufficiently long tails
for the solutions.

After constructing the appropriate initial potential $\Phi$, the mG equation was integrated using
a fourth-order Runge-Kutta method.
We used finite differences to approximate the spatial derivatives and applied periodic boundary conditions.
To avoid interference from the periodic boundary assumption, we had to choose a sufficiently large interval length.

The simulations reveal that the supersoliton breaks up during the collision, so that only regular solitons emerge
after the collision.
To illustrate this, we discuss a typical result obtained from simulations
with $D=-\sqrt{3.6}$.
The initial disturbance consists of a supersoliton with velocity
$v=0.5\left(v_{DL}+v_{\mathrm{max}}\right)\approx0.1218$
that is shifted $\eta_0=75$ units to the left, and a slower soliton with velocity $v=0.1$.
For this simulation, the interval length is $L=1200$ and a grid with $N=25600$ points are used.
Therefore, the spatial width is $\Delta\eta\approx0.047$.
An integration increment of $\Delta t=10^{-4}$ is used.

\begin{figure}
\begin{centering}
\includegraphics[scale=0.45]{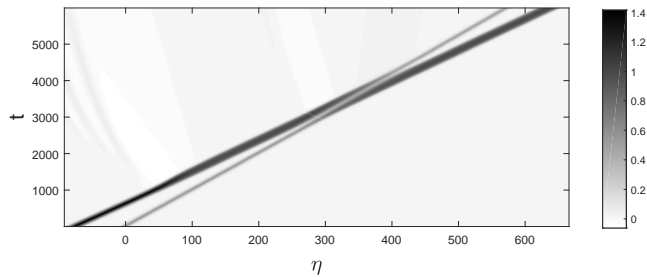}
\par\end{centering}
\caption{Simulation of a supersoliton overtaking a regular soliton.
The electrostatic potential $\Phi$ is plotted as a function of $\eta$ and $t$.}
\label{fig:Collision 1}
\end{figure}

The results are shown in Figure~\ref{fig:Collision 1}, where the magnitude of the solution $\Phi$ is plotted
as a function of $\eta$ and $t$.
Here we see that the supersoliton (initially on the left) widens around $t \approx 1000$, before the collision takes place.
The fact that the supersoliton breaks up during this time is not obvious from the figure.
The
amplitude of the
collision peaks around $t=3600$, before two solitons with smaller amplitudes emerge.
It is therefore clear that the collision is inelastic.

\begin{figure}
\includegraphics[scale=0.35]{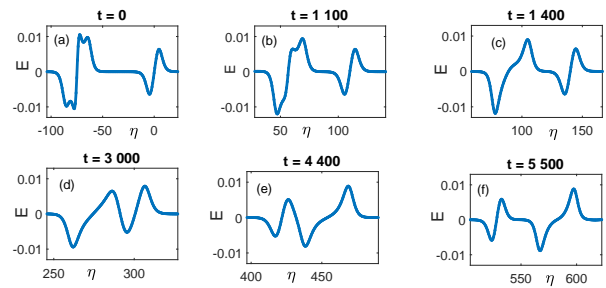}
\caption{Simulation of supersoliton overtaking a regular soliton.
The electric field $E$ is shown at different times $t$,
as specified on top of each panel.}
\label{fig:Collision 2}
\end{figure}

To see the breaking up of the supersoliton more clearly, in Figure~\ref{fig:Collision 2}
we graphed the $\eta$ profiles of the electric field, $E = -\partial\Phi/\partial\eta,$ at different
values of $t.$
In panel (a) of Figure~\ref{fig:Collision 2}, the initial condition is shown.
The characteristic ``wiggles" of the supersoliton are clearly visible to the left
of the regular soliton.
As the supersoliton approaches the regular soliton, the supersoliton starts to deform.
This is shown in panel (b) for $t=1100$.
The supersoliton breaks up to form a regular soliton.
Panel (c) shows the solution at $t=1400$, after the supersoliton deformed to become a regular soliton.

The collision of the resulting two solitons is shown in panels (d)--(f) of Figure~\ref{fig:Collision 2}.
The faster soliton overtakes the slower, resulting in a transient solution
as shown in panel (d) for $t=3000$.
Eventually, the faster soliton re-emerges in front of the slower one, as shown in panel (e) for $t=4400$.
Beyond $t=4400$, the separation between the two solitons increases, as depicted in panel (f) for $t=5500$.

\section{Conclusions and future work}

In this paper, we applied reductive perturbation analysis to study small-amplitude supersolitons in a plasma
consisting of cold ions and two-temperature Boltzmann electrons.
To do so, we considered near-supercritical plasma compositions.
We derived a generalized Korteweg-de Vries equation, referred to as the modified Gardner equation.
For that equation, we derived the necessary conditions for small-amplitude supersolitons to exist.

We also used the equation to study the collision properties of small-amplitude supersolitons,
both theoretically and through simulations.
Theoretically, we showed that in contrast to the KdV and mKdV equations, the mG equation is not
completely integrable.
Hence, collisions of small-amplitude supersolitons will be inelastic.
Numerical simulations of the collisions between solitons and supersolitons show
that the supersolitons break up during the collision to form a regular soliton.
This is very different from elastic collisions of regular solitons.

These results show that, in the small-amplitude regime, supersolitons are not as robust as regular solitons,
and may break up during collisions.
This suggests that their life spans may be much shorter than that of regular solitons
and might
explain the low number of supersoliton observations in space plasmas.

However, caution must be taken in the interpretation of these results.
Indeed, for these conclusions to be valid, our results must be extended beyond the small-amplitude regime.
To do so, one has to study the collision properties of
supersolitons in laboratory experiments or numerical simulations.
In addition, head-on collisions lie beyond the scope of this analysis.

In conclusion, we hope that our results will generate interest
in the topic of supersoliton collisions, and that this study can be used as a benchmark for
further investigations.

\section*{Acknowledgement}

CO wishes to acknowledge the financial assistance of the National Research Foundation (NRF)
towards this research. Opinions expressed and conclusions arrived at, are those of the authors
and are not necessarily to be attributed to the NRF.

\end{document}